 \documentclass[final,3p,times,twocolumn]{elsarticle}

\usepackage{graphicx}
\usepackage{amssymb}
\usepackage{amsmath}
\usepackage{units}
\usepackage{multirow}
\usepackage{bigdelim}
\usepackage{threeparttable}
\usepackage{lineno}

\biboptions{sort&compress}

\journal{Nuclear Instruments and Methods in Physics Research A}

\begin{document}
\bibliographystyle{elsarticle-num}

\begin{frontmatter}



\title{Lifetime measurement of excited low-spin states via the $(p,p^{\prime}\gamma$) reaction}


\author[col]{A.~Hennig\corref{cor1}}
\ead{hennig@ikp.uni-koeln.de}
\cortext[cor1]{Corresponding author}
\author[col]{V.~Derya}
\author[sofia,inrne]{M.~N.~Mineva}
\author[col,inrne]{P.~Petkov}
\author[col]{S.~G.~Pickstone}
\author[col]{M.~Spieker}
\author[col]{A.~Zilges}

\address[col]{Institute for Nuclear Physics, University of Cologne, Z\"ulpicher Stra{\ss}e 77, D-50937 Cologne, Germany}
\address[sofia]{Faculty of Physics, University of Sofia, BG-1164 Sofia, Bulgaria}
\address[inrne]{Institute for Nuclear Research and Nuclear Energy INRNE, Bulgarian Academy of Sciences, BG-1784 Sofia, Bulgaria}

\begin{abstract}
In this article a method for lifetime measurements in the sub-picosecond regime via the Doppler-shift attenuation method (DSAM) following the inelastic proton scattering reaction is presented. In a pioneering experiment we extracted the lifetimes of 30 excited low-spin states of $^{96}$Ru, taking advantage of the coincident detection of scattered protons and de-exciting $\gamma$-rays as well as the large number of particle and $\gamma$-ray detectors provided by the SONIC@HORUS setup at the University of Cologne. The large amount of new experimental data shows that this technique is suited for the measurement of lifetimes of excited low-spin states, especially for isotopes with a low isotopic abundance, where $(n,n^{\prime}\gamma$) or - in case of investigating dipole excitations - ($\gamma,\gamma^{\prime}$) experiments are not feasible due to the lack of sufficient isotopically enriched target material.

\end{abstract}

\begin{keyword}
$\gamma$-ray spectroscopy \sep Doppler-shift attenuation method \sep lifetime measurement \sep p$\gamma$-coincidence technique
\end{keyword}

\end{frontmatter}

\section{Introduction}
\label{sec:introduction}
Absolute electromagnetic transition strengths are valuable observables in nuclear structure physics since they are related to the nuclear wave functions of the initial and final states involved. They are proportional to the square of the reduced transition matrix elements and can be measured via, e.g., Coulomb-excitation \cite{Glas98} and ($\gamma,\gamma^{\prime}$) \cite{Knei06} experiments, directly. However, these techniques are selective to dipole and quadrupole excitations from the ground state. In a complementary approach, absolute transition probabilities can be determined if the spin and parity quantum numbers of the states involved, the decay branching ratio and the multipolarity of the $\gamma$ transition, as well as the lifetime $\tau$ of the initial state are known. Thus, the measurement of lifetimes of excited states is a key ingredient for the determination of absolute transition strengths. 

While for lifetimes in the nano- and picosecond regime fast-timing \cite{Mach89,Mosz89} and Recoil Distance Doppler-Shift methods (RDDM) \cite{Dewa12} can be effectively applied, the Doppler-shift attenuation method (DSAM) \cite{Nola79, Alex78} is usually the method of choice for the measurement of sub-picosecond lifetimes. The latter utilizes the Doppler shift of $\gamma$-rays emitted during the slowing-down process from recoiling target nuclei after a nuclear reaction and is described in detail in, e.g., Refs.~\cite{Nola79, Alex78}. Usually, a nuclear reaction induced by a heavy-ion beam is favored to ensure a large momentum transfer, resulting in a high recoil velocity and thus a large Doppler-shift of the measured $\gamma$-ray energies in order to use the details of the lineshape for the subsequent data analysis \cite{Petk98}. However, reactions induced by heavy-ion beams generally favor the population of high-spin states. 

In contrast, comprehensive lifetime data for low-spin states have been obtained using the Doppler-shift attenuation technique following the inelastic neutron-scattering reaction (DSAM-INS), developed and extensively used at the University of Kentucky, see, e.g., Ref.~\cite{Belg96}. Because of the reduced recoil velocity, the centroid-shift method was applied instead of a lineshape analysis. It exploits the angular dependence of the centroid of the $\gamma$-ray energies

\begin{equation}
E_{\gamma}(\Theta)=E_{\gamma}^0\left(1+F(\tau)\frac{v_0}{c}\cos\Theta\right)
\label{eq:shift}
\end{equation} 

to extract the Doppler-shift attenuation factor $F(\tau)$ \cite{Nola79}. Here, $E_{\gamma}^0$ denotes the unshifted $\gamma$-ray energy, $\Theta$ is the angle between the direction of motion of the reaction product and the direction of $\gamma$-ray emission, and $v_0$ denotes the initial recoil velocity. The lifetime of the excited state is finally obtained by comparing the experimental Doppler-shift attenuation factor with a calculated one, based on the modeling of the slowing-down process of the recoil nucleus in the target material.

A crucial limitation of the DSAM-INS method is the large amount of about a few tens of gram of isotopically enriched target material which is usually needed in the $(n,n^{\prime}\gamma$) reaction. Consequently, this method is not feasible for the investigation of isotopes with a low isotopic abundance since a suitable target is too expensive in most cases. 

This limitation can be overcome by applying the inelastic proton-scattering reaction which typically requires about 0.5~$\mathrm{mg}$ of target material - \textit{thus about a factor of $10^5$ less material compared to the amount needed for $(n,n^{\prime}\gamma$) experiments.} Furthermore, the proton in the exit channel of the $(p,p^{\prime}\gamma$) reaction can be detected with charged-particle detectors that can be placed close to the target. The coincident detection of the scattered protons and the de-exciting $\gamma$-rays has several additional advantages \cite{Enge69, Seam69}:
\begin{itemize}
\item The direction and the initial velocity of the recoiling target nuclei are precisely defined by the reaction kinematics, which reduces the uncertainty of the angle $\Theta$ that governs the Doppler shift (see Eq.~(\ref{eq:shift})).
\item The excitation energy of the recoil nucleus can be extracted from the energy information of the scattered proton on an event-by-event basis. Thus, the peak centroids can be extracted from $\gamma$-ray spectra that were gated on the excitation energy of the level of interest in order to eliminate feeding contributions from higher-lying states. Furthermore, the proton-gated spectra are characterized by an increased peak-to-background ratio, so that even weak transitions can be taken into account for the data analysis.
\item If the charged-particle detectors are positioned at predominantly backward angles, events with large momentum transfer to the recoiling target nuclei are selected. The enhanced recoil velocity results in a larger Doppler shift of the emitted $\gamma$-rays.
\end{itemize}

This paper is devoted to the method of lifetime measurement utilizing the $(p,p^{\prime}\gamma$) reaction and benefits from the coincident detection of scattered protons and de-exciting $\gamma$-rays. Though this method has been exploited already some decades ago \cite{Enge69, Seam69}, those experiments were suffering from low statistics due to the coincidence requirement and the small number of detectors used \cite{Alex78}. In the present experiment on $^{96}$Ru, 14~HPGe detectors of a $\gamma$-ray spectrometer were used in combination with six detectors of a particle-detector array. Taking advantage of the high p$\gamma$-coincidence efficiency achieved with the large amount of detectors, the lifetimes of 30 excited states of one nucleus could be extracted from a single experiment, proving this method to be very powerful for lifetime measurements. The nuclear structure physics interpretation has been published separately \cite{Henn14a} and will not be subject of the present article.

\section{Experimental setup}

\begin{table}[t!]
\begin{center}
\caption{Details of the particle detectors of the SONIC arrary. The table shows the positions, i.e., the angle with respect to the beam axis $\theta_{p^{\prime}}$ and the azimutal angle $\phi_{p^{\prime}}$ of the six PIPS detectors used in the present experiment (see Fig. \ref{fig:dsam_geo} for the definition of $\theta_{p^{\prime}}$ and $\phi_{p^{\prime}}$). The last two columns shows the target-to-detector distance $d$ for each detector and the corresponding solid angle coverage $\Omega$, respectively.} \label{tab:setup}\vspace{2mm}
\begin{threeparttable}
\begin{tabular}{p{0.5cm}p{1.3cm}p{1.3cm}p{1.3cm}p{1.3cm}}
\hline\hline
Nr.				&	$\theta_{p^{\prime}}$ [$^{\circ}$]	&	$\phi_{p^{\prime}}$	[$^{\circ}$] &	d [mm]	& $\Omega$ [\%]\tnote{a} \\
\hline
1	&	131(5)	&	217(5)	&	46.5(10)	&	0.55(2) \\
2	&	131(5)	&	323(5)	&	46.5(10)	&	0.55(2)\\
3	&	122(5)	&	123(5)	&	46.5(10)	&	0.55(2)\\
4	&	122(5)	&	57(5)	&	46.5(10)	&	0.55(2)\\
5	&	61(5)	&	227(5)	&	100(1)		&	0.12(2)\\
6	&	61(5)	&	313(5)	&	100(1)		&	0.12(2)\\
\hline\hline
\end{tabular}
\begin{tablenotes}
        \item[a] relative to $4\pi$
    \end{tablenotes}
\end{threeparttable}
\end{center}
\end{table}

A $7.0~\mathrm{MeV}$ proton beam, provided by the 10~MV FN Tandem accelerator at the Institute for Nuclear Physics of the University of Cologne, impinged on a $106~\mathrm{\mu g/cm^2}$ enriched $^{96}$Ru target. The beam energy corresponded to a maximum recoil velocity of $v_0^{\mathrm{max}}/c = 2.5\cdot 10^{-3}$. The target was supported by a $^{12}$C backing with a thickness of $14~\mathrm{\mu g/cm^2}$, acting as a stopper material for the recoiling ions. The thicknesses and compositions of the target and stopper materials were confirmed via a Rutherford Backscattering Spectrometry (RBS) measurement prior to the experiment. 

For the $\gamma$-ray detection the HORUS spectrometer was used, housing 12 single-crystal HPGe detectors and two Clover-type detectors, positioned at angles of 35$^{\circ}$ (2x), 45$^{\circ}$ (2x), 90$^{\circ}$ (6x), 135$^{\circ}$ (2x), and 145$^{\circ}$ (2x) with respect to the beam axis (see, e.g., Ref. \cite{Nett14} for details on the HORUS spectrometer). With the target-to-detector distances used in the present experiments the solid angle coverage of the HPGe detectors was between 1.0~\% and 1.5~\% relative to a full solid angle coverage of $4\pi$. The Clover-type detectors as well as four single-crystal HPGe detectors were equipped with BGO-shields for an active Compton suppression. 

The scattered protons were detected with the particle-detector array SONIC, housing six passivated implanted planar silicon (PIPS) detectors with a thickness of $1.5~\mathrm{mm}$ and an active area of $A=150~\mathrm{mm^2}$, each. Because of the close geometry of the setup, they were positioned in the spacings between the HPGe detectors of the HORUS array, resulting in angles of 131$^{\circ}$ (2x), 122$^{\circ}$ (2x), and 61$^{\circ}$ (2x) with respect to the beam axis (see Tab. \ref{tab:setup} for details). An energy resolution of about $15~\mathrm{keV}$ was achieved for the particle detectors using calibration sources, while the energy resolution degrades to about $70~\mathrm{keV}$ in the in-beam experiment because of straggling in the target and stopper material as well as kinematic effects.

\begin{figure}[t!]
	\begin{center}
	\includegraphics[width=0.47\textwidth]{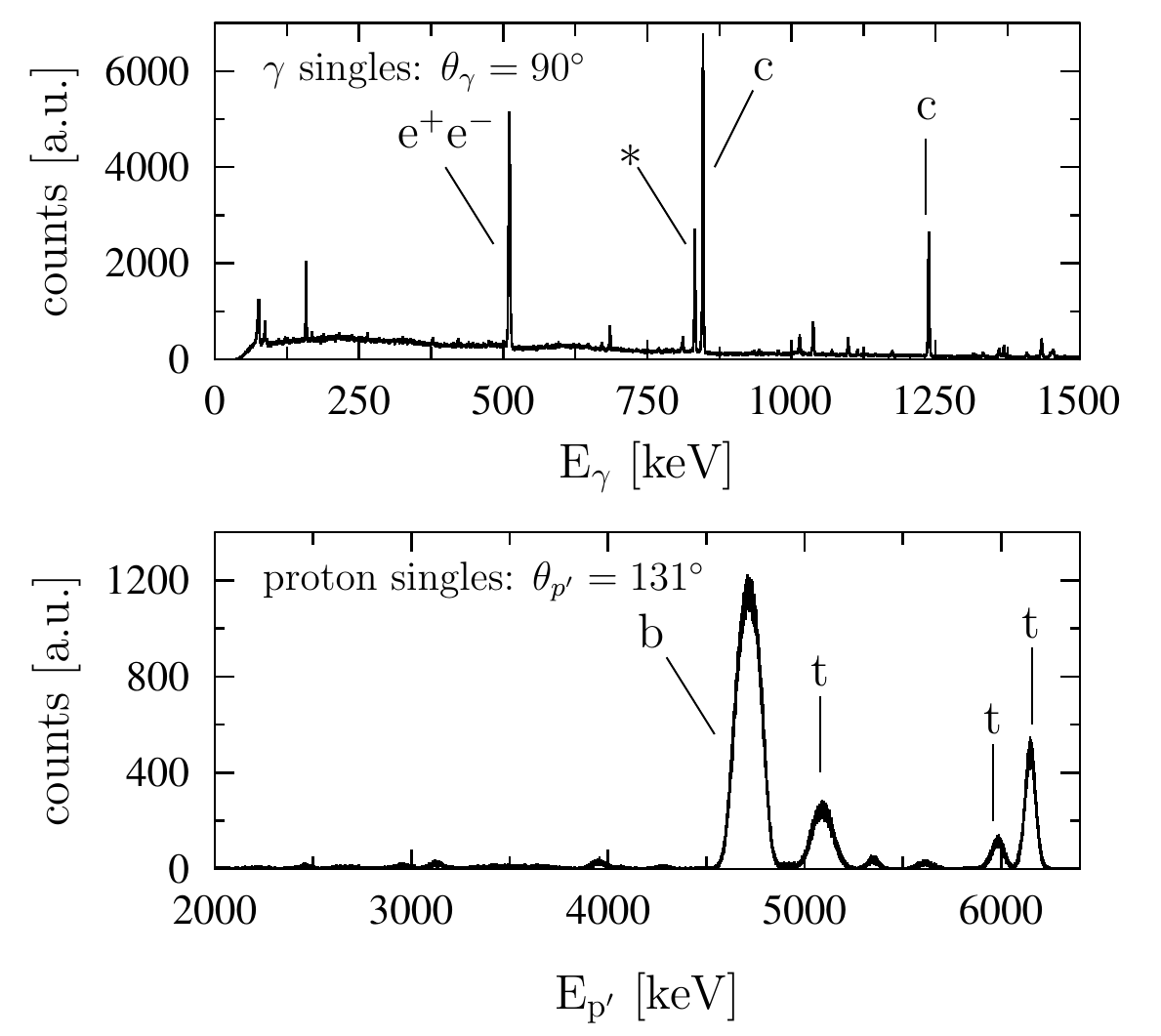}
	\end{center}
\caption{Typical $\gamma$-ray singles spectrum (upper panel) and proton singles spectrum (lower panel) obtained in the $^{96}$Ru$(p,p^{\prime}\gamma$) experiment at angles of $\theta_{\gamma}=90^{\circ}$ and $\theta_{p^{\prime}}=131^{\circ}$, respectively. The former is dominated by the $2^+_1 \rightarrow 0^+_{\mathrm{g.s}}$ $\gamma$-ray transition of $^{96}$Ru (indicated with $^{\ast}$) and transitions stemming from the $^{56}$Co calibration source (c). The proton spectrum is dominated by protons scattered in the target and backing material, indicated with (t) and (b), respectively.}
\label{fig:spectra}
\end{figure}

The detector signals were processed with the digital data acquisition modules DGF-4C Rev. F from the company XIA \cite{Hubb99} generating listmode data that contained the energy and time information of the active detector channels. The high spectral resolution of 32k in the energy spectra provided by the DGF-4C modules was mandatory to extract centroid shifts in the sub-keV regime. In total, data were acquired for 174 hours at an average beam current of $150~\mathrm{nA}$. Typical proton and particle spectra are shown in Fig.~\ref{fig:spectra}.

To enable a run-by-run energy calibration for each HPGe detector, a $^{56}$Co calibration source was mounted on the target ladder throughout the entire experiment. With this, a precision of $\pm 0.05~\mathrm{keV}$ was achieved for the energy calibration of the HPGe detectors, which is sufficient to extract centroid shifts in the sub-keV regime. 

\section{Data analysis}

\subsection{Reaction kinematics}
\begin{figure}[t!]
	\begin{center}
	\includegraphics[width=0.47\textwidth]{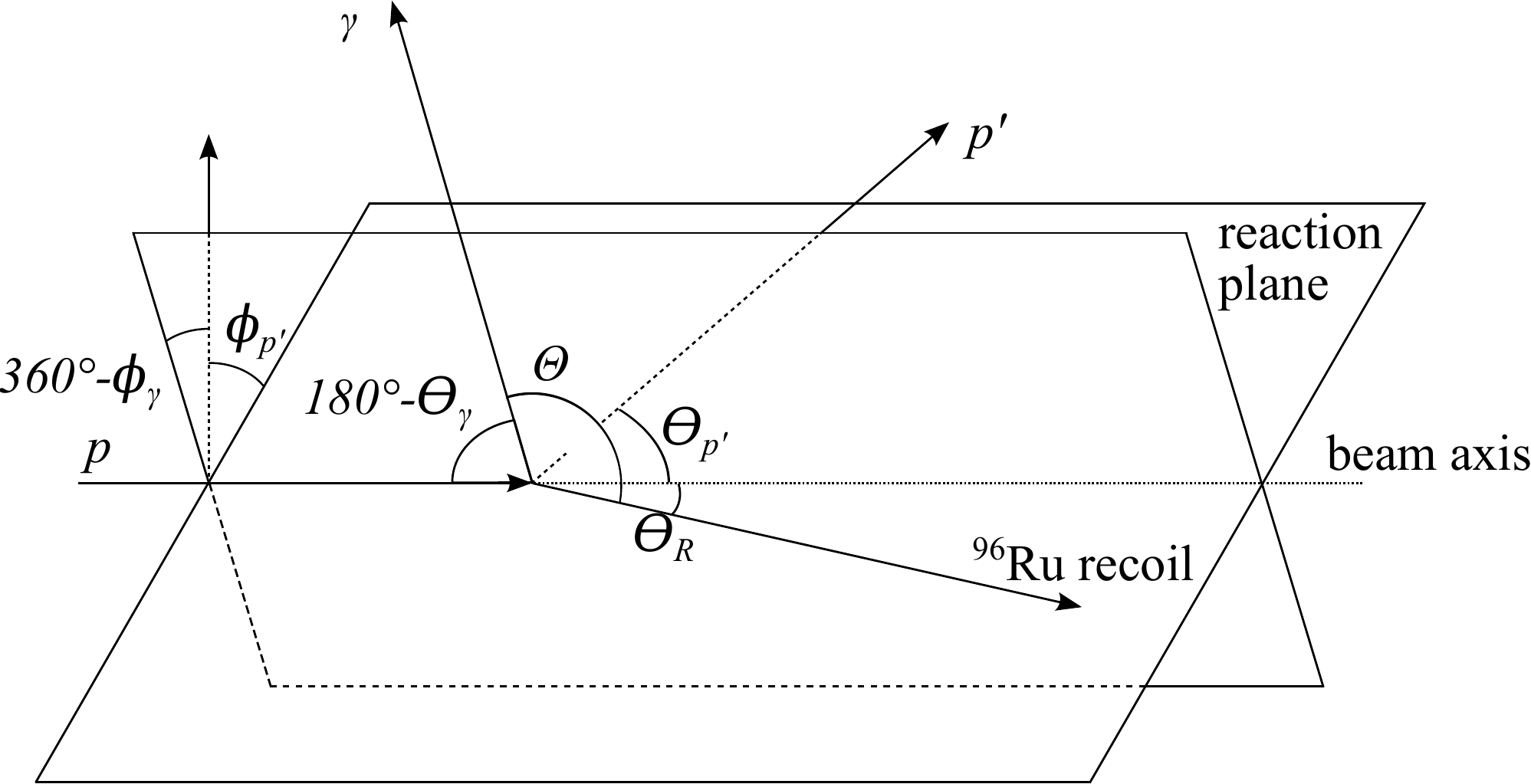}
	\end{center}
\caption{Kinematics of the $^{96}$Ru$(p,p^{\prime}\gamma$) reaction. Due to the conservation of linear momentum the reaction proceeds in a plane. The angle $\Theta$ between the direction of motion of the recoil nucleus and the direction of $\gamma$-ray emission governs the observed Doppler shift.}
\label{fig:dsam_geo}
\end{figure}

Figure \ref{fig:dsam_geo} shows the kinematics in the $^{96}$Ru$(p,p^{\prime}\gamma$) reaction. $\theta_{p^{\prime}}$ and $\theta_{R}$ denote the scattering angles of the proton and the recoil nucleus with respect to the beam axis, respectively, while $\phi_{p^{\prime}}$ is the azimuthal angle of the scattered proton in the laboratory frame. If the de-exciting $\gamma$-ray is detected in an HPGe detector placed at an angle of $\theta_{\gamma}$ with respect to the beam axis and at an azimuthal angle of $\phi_{\gamma}$, the angle $\Theta$ between the recoil nucleus and the $\gamma$-ray can be expressed as

\begin{equation}
\cos\Theta = \frac{1}{\sqrt{1+x^2}}\left[\cos\theta_{\gamma}-x\cdot\cos\left(\phi_{\gamma}-\phi_{p^{\prime}}\right)\right],
\label{eq:cos_theta}
\end{equation}

with 

\begin{equation}
x\equiv\tan\theta_R = \frac{\sin\theta_{p^{\prime}}}{\sqrt{\frac{E_p}{E_{p^{\prime}}}}-\cos\theta_{p^{\prime}}}~.
\end{equation}

\begin{figure}[t!]
	\begin{center}
	\includegraphics[width=0.47\textwidth]{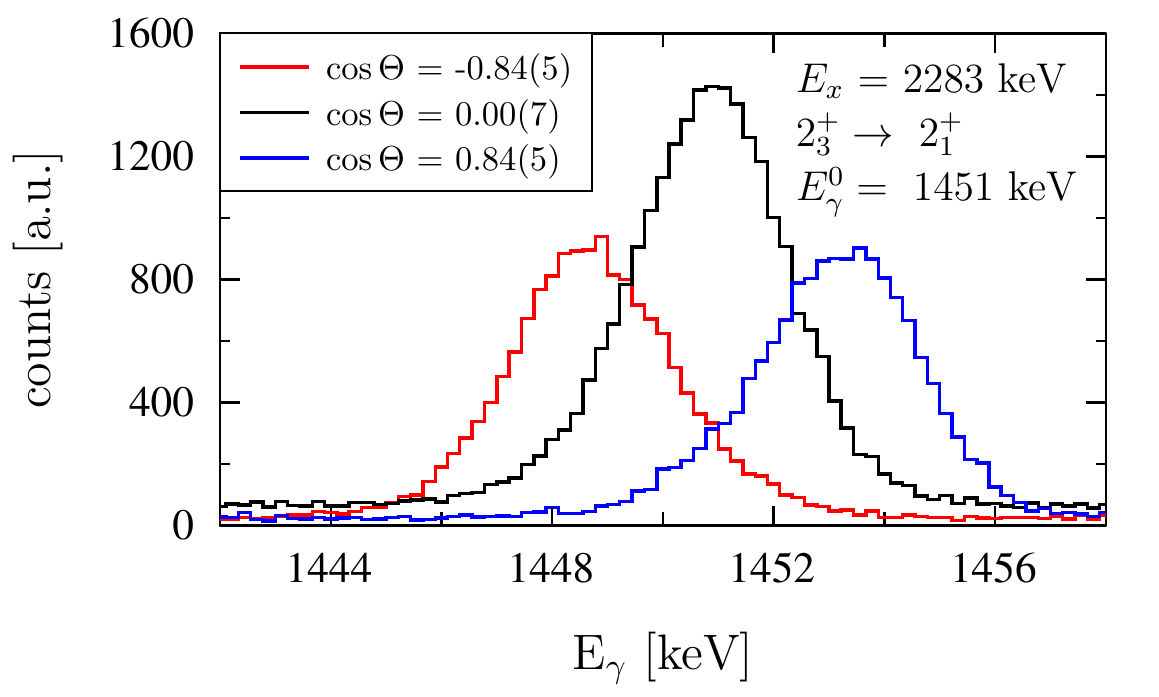}
	\end{center}
\caption{(Color online) $\gamma$-ray spectra gated on the excitation energy of the $2^+_3$ state at $E_x=2283~\mathrm{keV}$, obtained for three different groups which are characterized by three different values of $\cos\Theta$. An unshifted peak is observed for $\cos\Theta = 0.00(7)$, while the peaks for $\cos\Theta = -0.84(5)$ and $\cos\Theta = +0.84(5)$ are shifted to lower and higher energies, respectively.}
\label{fig:shifts}
\end{figure}

Here, $E_p$ and $E_{p^{\prime}}$ denote the energy of the incident and scattered proton. The energy of the scattered proton can be related to the excitation energy of the recoil nucleus via 

\begin{eqnarray}
E_x & =&  E_p\left(1-\frac{m_p}{m_t}\right)-E_{p^{\prime}}\left(1+\frac{m_p}{m_t}\right)+ \nonumber \\
	& & 2\frac{m_p}{m_t}\sqrt{E_{p}E_{p^{\prime}}}\cos\theta_{p^{\prime}}-E_{\mathrm{loss}}~,
\label{eq:ex_energy}	
\end{eqnarray}

\begin{figure*}[t!]
	\begin{center}
	\includegraphics[width=0.95\textwidth]{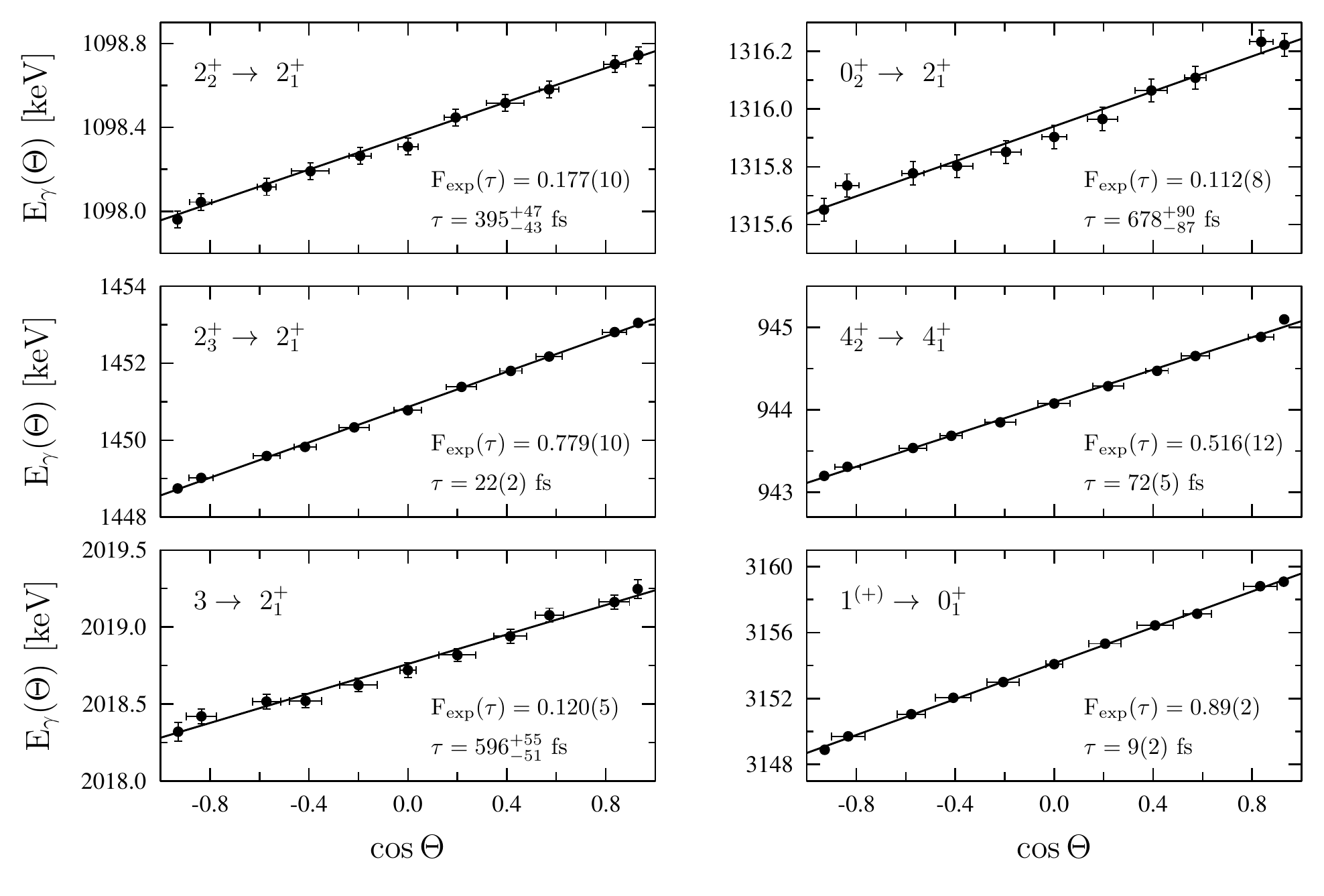}
	\end{center}
\caption{Centroid shifts of the peaks observed in the proton-gated $\gamma$-ray spectra as a function of the $\cos\Theta$ value characterizing the groups. The linear trend of the peak centroids is clearly visible. The error bars in horizontal direction include the standard deviation to the mean value of $\cos\Theta$ for each group as well as the uncertainty introduced by the opening angles of the particle detectors. The uncertainties of the peak centroids are dominated by the uncertainty in the energy calibration with the $^{56}$Co source.}
\label{fig:centroids}
\end{figure*}

where $m_p$ and $m_t$ denote the masses of target and projectile, while $E_{\mathrm{loss}}$ is the energy loss of the scattered protons in the target and stopper material \cite{Catf86}. Thus, Eq.~(\ref{eq:cos_theta}) is a function of the beam energy $E_p$, the excitation energy $E_x$, and the positions of the silicon and HPGe detectors only.

\subsection{Data sorting}

For a fixed excitation and beam energy, there are 84 silicon vs. HPGe detector pairs, each pair with a different value of $\cos\Theta$. These pairs were sorted to eleven groups, such that the difference for $\cos\Theta$ is less than 0.2 for each pair in every group. With this, each group was characterized by the average value of $\cos\Theta$ of all pairs within the group. From the acquired proton-$\gamma$ coincidence data eleven coincidence matrices were generated by adding up the data of all silicon vs. HPGe detector pairs within each group. According to Eq.~(\ref{eq:cos_theta}), the assignment of detector pairs to the different groups depends on the excitation energy under consideration. Note that the chosen ``binning'' of 0.2 and the resulting number of eleven groups was chosen to provide a compromise between sufficient statistics in the proton-gated $\gamma$-ray spectra of each group and angular granularity for the centroid shift.

To experimentally extract the Doppler-shift attenuation factor $F{_\mathrm{exp}}(\tau)$, $\gamma$-ray spectra were generated for each group by gating on the excitation energy of interest. Figure \ref{fig:shifts} shows the proton-gated $\gamma$-ray spectra for three groups characterized by different values of $\cos\Theta$ in the energy region of the $2^+_3\rightarrow 2^+_1$ transition at $E_{\gamma}^0 =1451~\mathrm{keV}$. The centroid shift is clearly visible: An unshifted peak is observed for the group characterized by an angle of $\cos\Theta = 0.00(7)$, while the peak is shifted to higher and lower energies for the groups characterized by angles of $\cos\Theta = 0.84(5)$ and $\cos\Theta = -0.84(5)$, respectively.

The centroid energy as a function of the $\cos\Theta$ value of each group is shown in Fig.~\ref{fig:centroids} for several $\gamma$-ray transitions in the nucleus $^{96}$Ru. The Doppler-shift attenuation factor $F{_\mathrm{exp}}(\tau)$ was extracted from a linear fit to the data according to Eq.~(\ref{eq:shift}). Note that different values of the initial recoil velocity $v_0$ had to be taken into account for the different particle detector positions and excitation energies under consideration.

\begin{figure}[t!]
	\begin{center}
	\includegraphics[width=0.47\textwidth]{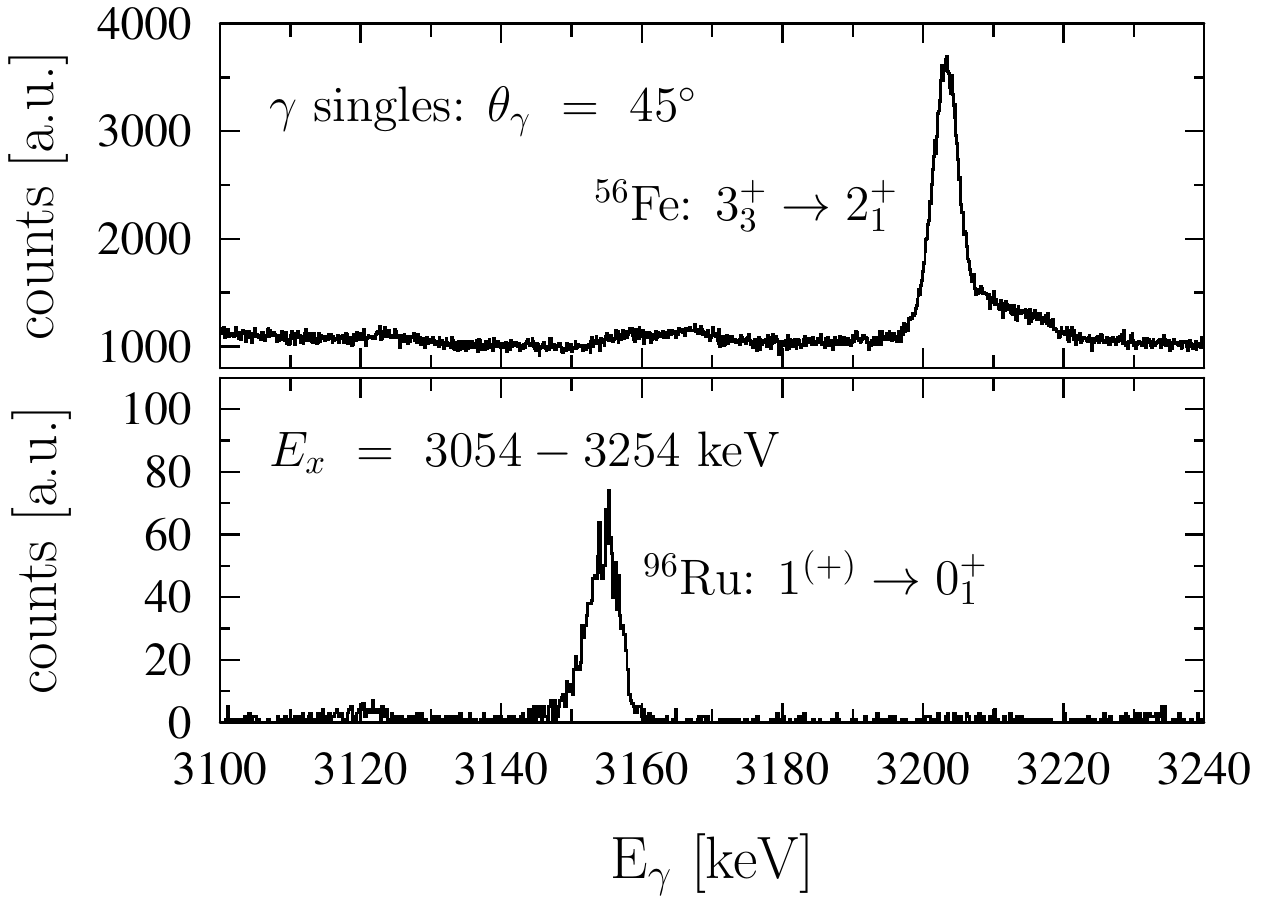}
	\end{center}
\caption{Singles $\gamma$-ray spectrum (upper panel) for an HPGe detector positioned at an angle of $45^{\circ}$ with respect to the beam axis and a proton-gated $\gamma$-ray spectrum of the group characterized by an angle of $\cos\Theta = 0.00(7)$ (lower panel). Due to the enhanced peak-to-background ratio the $1^{(+)}\rightarrow 0^+$ transition of $^{96}$Ru can be resolved only in the proton-gated spectrum.}
\label{fig:gated_DSAM}
\end{figure}

The need for the proton-$\gamma$ coincidence technique to resolve weak transitions from the background is illustrated in Fig.~\ref{fig:gated_DSAM}. While in the singles HPGe spectrum the peak at $E_{\gamma}^0=3154~\mathrm{keV}$ is hidden in the background, it can be clearly resolved in the proton-gated spectrum and thus be analyzed only with the coincidence requirement.

\subsection{Theoretical description of the slowing-down process}
Modeling the slowing-down process of the recoil nucleus in the target and stopper material yields a relation between the Doppler-shift attenuation factor and the nuclear level lifetime. A comparison with the experimental value $F{_\mathrm{exp}}(\tau)$ finally yields the experimental lifetime value. The description of the stopping process follows the approach presented in Ref.~\cite{Petk98} and will be sketched only briefly in the following.

According to the theory of Lindhard, Scharff and Schi\o tt (LSS) \cite{Lind63}, the energy loss of an ion traversing a medium can be written as the sum of an electronic and a nuclear part

\begin{equation}
\frac{d\varepsilon}{d\rho}=\left(\frac{d\varepsilon}{d\rho}\right)_{\mathrm{e}}+f_n\left(\frac{d\varepsilon}{d\rho}\right)_{\mathrm{n}}~,
\end{equation}

\begin{figure}[t!]
	\begin{center}
	\includegraphics[width=0.47\textwidth]{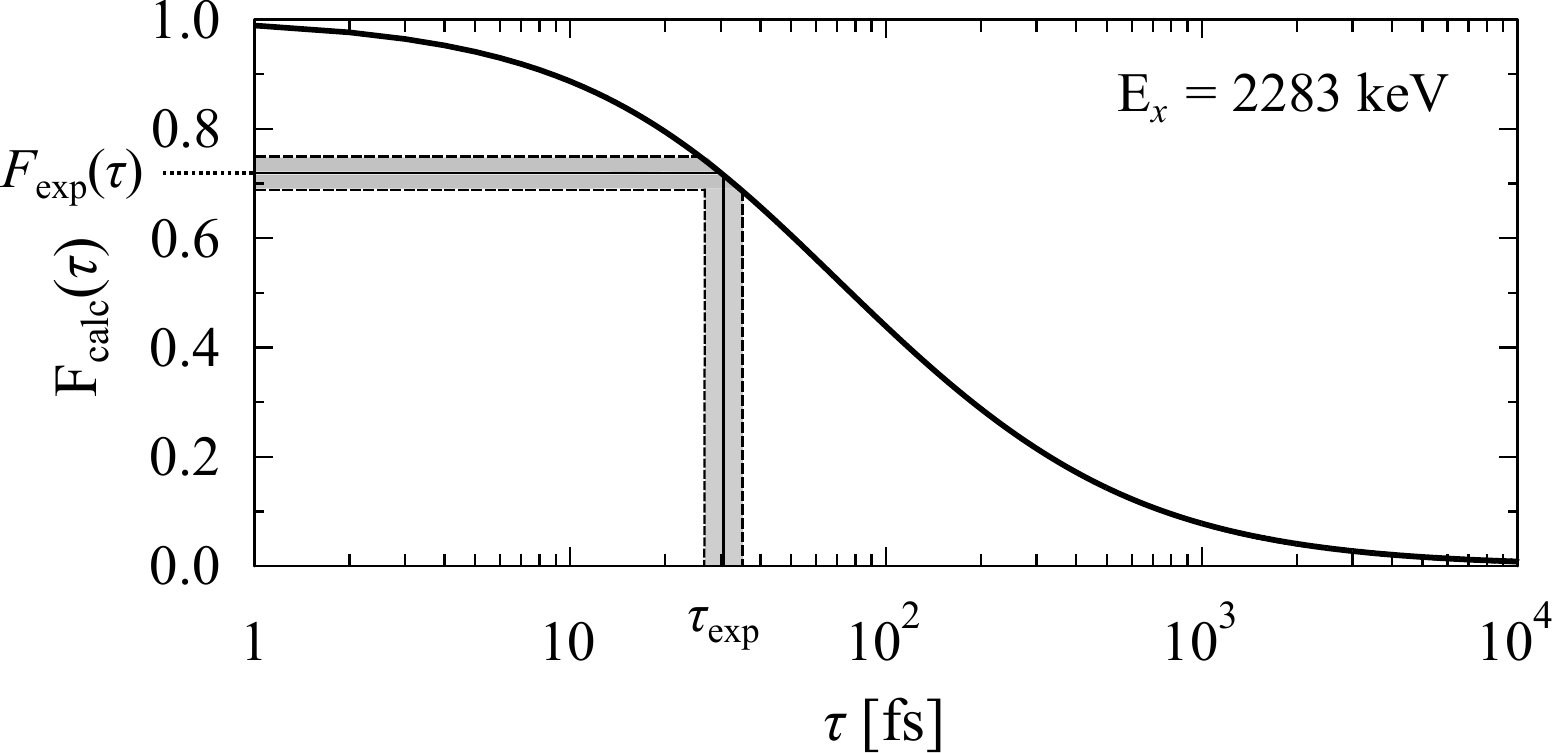}
	\end{center}
\caption{Calculated Doppler-shift attenuation factor $F_{\mathrm{calc}}(\tau)$ for the $2^+_3$ state at $E_x = 2283~\mathrm{keV}$ as a function of the level lifetime $\tau$. The experimental lifetime value $\tau_{\mathrm{exp}}$ is obtained by projecting the experimental Doppler-shift attenuation factor $F_{\mathrm{exp}}(\tau)$ to the x-axis. The gray-shaded area corresponds to the statistical uncertainties.}
\label{fig:ftau}
\end{figure}

with the dimensionless variables $\varepsilon$ and $\rho$ for the ion energy and distance. The electronic stopping power was parametrized according to the generalized LSS formula \cite{Petk98} and its parameters were determined by using the tabulated stopping power values from the semi-empirical tables of Northcliffe and Schilling (NS) \cite{Nort70}, corrected for the electronic structures according to Ziegler and Chu \cite{Zieg74}. The correction factor $f_n$ for the nuclear stopping accounts for deviations of the description with a Thomas-Fermi potential. A value of $f_n=0.7$ has been found empirically for various combinations of recoil ions and stopper materials \cite{Kein85} and thus, this value was adopted for the present calculations. The stopping process is finally described with the Monte-Carlo simulation code \textsc{dstop96} \cite{Petk98} which is based on the code \textsc{desastop} \cite{Wint83a, Wint83b} and follows the approach of Ref.~\cite{Curr69}. In the code, the finite opening angles of the silicon and HPGe detectors are explicitly taken into account to obtain the calculated Doppler-shift attenuation factor $F_{\mathrm{calc}}(\tau)$. In Fig.~\ref{fig:ftau}, the latter is shown as a function of the level lifetime $\tau$. Note again, that the calculation depends on the initial recoil velocity and thus on the excitation energy of the level of interest.

\section{Results}

\begin{table*}[t!]
\begin{center}
\caption{Results of the lifetime measurement of excited states of $^{96}$Ru via the $(p,p^{\prime}\gamma$) reaction for states where previously measured lifetime values are reported in the literature. In column one and two, the excitation energy and the spin and parity quantum numbers of the excited state are shown. The energy of the $\gamma$-ray transition used for the DSAM analysis is shown in column three, while the extracted Doppler-shift attenuation factor is given in column four. The resulting lifetime value extracted in this work is presented in column five. For the state at $E_x=2283~\mathrm{keV}$ two depopulating transitions were used to extract $F_{\mathrm{exp}}(\tau)$ and the weighted average is quoted in column six. Previously measured lifetime values are given in column eight for comparison. For the dipole transitions, the correction factor for the newly observed branching ratios is given in column seven (see text for details).} \label{tab:results}\vspace{2mm}

\begin{tabular*}{\textwidth}{p{1.7cm}p{0.7cm}p{2.0cm}p{2.0cm}p{1.5cm}p{1.5cm}p{1.5cm}p{2.2cm}}
\hline\hline
$E_x$ [keV]				&	$J^{\pi}$	&	$E_{\gamma}$ [keV]	&	$F_{\mathrm{exp}}(\tau)$	&	$\tau$ [fs]	&	$\tau_\mathrm{av}$ [fs]	&	$\sum\limits_f\frac{\Gamma_f}{\Gamma_0}$	&	$\tau_{\mathrm{lit}}$ [fs]			\\

\hline

1930.9(2)			& $2^+$			& 1098.4(1)			& 0.177(10)		&	$395^{+47}_{-43}$	&	&			&	$550^{+220}_{-160}$	\cite{Adam86} \\

2148.5(2)			& $0^+$ 		& 1315.9(1)			& 0.112(8)		&	$678^{+90}_{-87}$	&	&		&	$660^{+910}_{-260}$	\cite{Adam86} \\
	
2283.3(2)			& $2^+$ 		& 1450.9(1)			& 0.779(10)		&	22(2)				&	\rdelim\}{2}{1.8cm}[\normalfont\hspace{2pt} 23(2)] &	&  \multirow{2}{2.2cm}{22(7) \cite{Piet01}}  \\	
					&				& 2283.3(1)			& 0.74(3)		&	27(4)				&	&									\\
			
2462.3(2)			& $4^+$			& 944.1(1)			& 0.516(12)		&	72(5)				&	&		&	$140^{+140}_{-70}$	\cite{Adam86} \\

2851.4(2)			& $3$			& 2018.8(1)			& 0.120(5)		&	$596^{+55}_{-51}$	&	&		&	$200^{+140}_{-70}$ \cite{Adam86} \\

3154.3(2)			& $1^{(+)}$		& 3154.1(1)			& 0.89(2)		&	9(2)				&	&	1.00(12)	&	5(2) \cite{Linn05a}		\\

3282.4(2)			& $1$			& 3282.2(1)			& 0.66(3)		&	37(6)				&	&	1.51(14)	&	47(7) \cite{Linn05a}	\\

3479.4(2)			& 1				& 3479.3(1)			& 0.61(3)		&	47(6)				&	&	1.14(12)	&	45(6) \cite{Linn05a}	\\

\hline\hline
\end{tabular*}
\end{center}
\end{table*}

The final lifetime values were extracted by comparing the experimental Doppler-shift attenuation factor to the theoretically predicted $F_{\mathrm{calc}}(\tau)$ curve. In addition to the statistical error, a systematic error was assigned to account for the imprecise knowledge of the stopping powers by varying the electronic stopping power by $10~\%$ in both directions. Similarly, a decrease/increase of the reduction factor of the nuclear stopping power $f_n$ by $10~\%$ would result in an increase/decrease of the obtained lifetime value which is of the same order of magnitude. The effect is slightly larger in case of a reduction of the nuclear stopping power and increases with increasing lifetime. The observed effect of a variation of the stopping powers is in agreement with the results obtained in Ref.~\cite{Petk98}.

For the state at $E_x=2283~\mathrm{keV}$ two de-exciting transitions were used for the lifetime analysis. This results in two independent values for $F_{\mathrm{exp}}(\tau)$. However, the resulting lifetime values are consistent within the uncertainties. Thus, the weighted average $\tau_{\mathrm{av}}$ is calculated to obtain the final result. We like to stress again, that the derived lifetimes are independent of feeding contributions due to the gate on the excitation energy of the level of interest. 

\begin{figure}[t!]
	\begin{center}
	\includegraphics[width=0.47\textwidth]{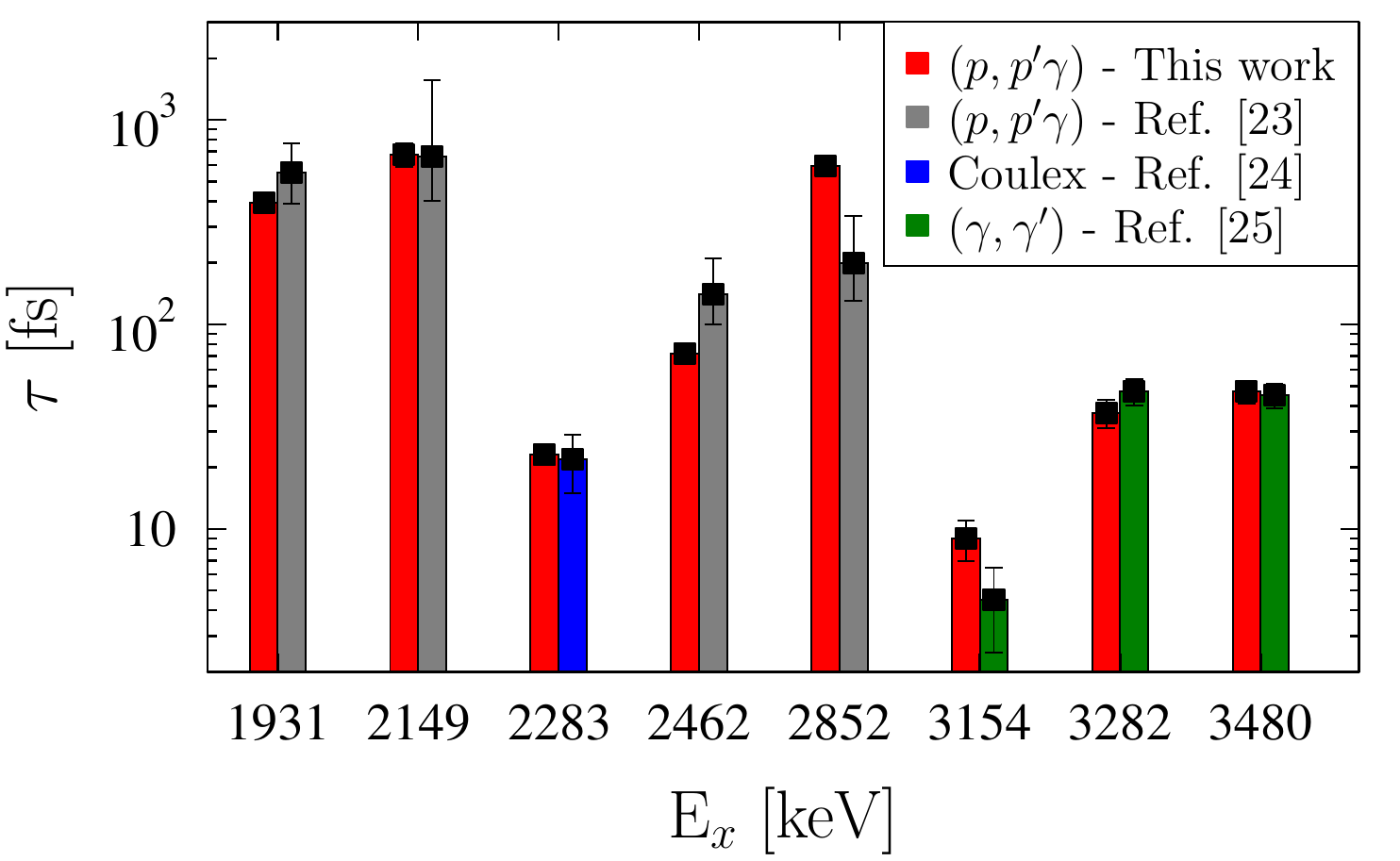}
	\end{center}
\caption{(Color online) Comparison of the lifetimes extracted in this work (red bars) compared to previous measurements. Especially if compared to the ($\gamma,\gamma^{\prime}$) data \cite{Linn05a} (green bars) and the Coulex data \cite{Piet01} (blue bar), the new data are in excellent agreement. Much lower experimental uncertainties are obtained compared to the $^{96}$Ru$(p,p^{\prime}\gamma$) data of Ref.~\cite{Adam86} (gray bars).}
\label{fig:known_taus}
\end{figure}

To validate the experimental procedure presented in this article, the lifetimes obtained in this work are compared to previously measured values. Lifetime values in the sub-picosecond regime for excited states in $^{96}$Ru have been reported from a $^{96}$Ru$(p,p^{\prime}\gamma$) experiment \cite{Adam86}, a ($\gamma,\gamma^{\prime}$) experiment \cite{Linn05a} and a Coulomb-excitation (Coulex) experiment in inverse kinematics \cite{Piet01}. The lifetime values obtained in this work are shown in comparison with the previously measured values in Tab.~\ref{tab:results}. A graphical comparison is shown in Fig.~\ref{fig:known_taus} as well. 

The agreement of the present results with the previous ones, especially for those obtained in the Coulex and the ($\gamma,\gamma^{\prime}$) experiment, is remarkable. Note that in the $^{96}$Ru($\gamma,\gamma^{\prime}$) experiment of Ref.~\cite{Linn05a} only the ground-state decays of the dipole transitions have been observed for all but the state at $E_x=3154~\mathrm{keV}$. Thus, the lifetime data reported in Ref.~\cite{Linn05a} are upper limits only. To be able to compare the reported values to the present data, the lifetime values of Ref.~\cite{Linn05a} were corrected for the $\gamma$-decay branching ratios obtained in the present experiment according to 

\begin{equation}
\tau = \frac{\tau^{\prime}}{\sum\limits_f\frac{\Gamma_f}{\Gamma_0}}~.
\end{equation} 

Here, $\tau^{\prime}$ is the lifetime value reported in Ref.~\cite{Linn05a} and $\frac{\Gamma_f}{\Gamma_0}$ the $\gamma$-decay branching ratio to the final state $f$ relative to the ground-state transition, i.e., $\Gamma_0$ corresponds to the $\gamma$-decay width to the ground state. For the dipole transitions, the correction factor $\frac{\Gamma_f}{\Gamma_0}$ is quoted in Tab.~\ref{tab:results} as well.

Except for the state at $E_x=2149~\mathrm{keV}$, some discrepancies between the present data and the ones reported in Ref.~\cite{Adam86} were obtained. The latter were measured via the DSAM technique following the $(p,p^{\prime}\gamma$) reaction as well. However, their values are characterized by large uncertainties, stemming predominantly from the extraction of the experimental Doppler-shift attenuation factor $F_{\mathrm{exp}}(\tau)$ (see especially Fig.~5 in Ref.~\cite{Adam86}). From the present experiment, the experimental uncertainties were reduced by at least a factor of two. Furthermore, the measurement in Ref.~\cite{Adam86} was carried out without additional particle detectors. Thus, their results might not be independent of feeding contributions and represent only upper limits for the lifetime of the excited states. Indeed, our newly measured lifetime values for the states at  $E_x=1931~\mathrm{keV}$ and  $E_x=2462~\mathrm{keV}$ are smaller by about a factor of two, pointing towards considerable feeding contributions for these states.

\section{Summary}
In this article, we presented a method of lifetime measurement via the $(p,p^{\prime}\gamma$) reaction. The method utilizes the centroid-shift version of the DSAM technique and profits from the coincident detection of scattered protons and de-exciting $\gamma$-rays. Unlike previous experiments of this kind \cite{Enge69, Seam69}, the measurement is not suffering from low statistics due to the coincidence requirement because of the increased amount of particle and $\gamma$-ray detectors provided by the SONIC@HORUS setup at the University of Cologne. The gating condition ensures the elimination of feeding contributions and highly increases the peak-to-background ratio in the spectra used for the extraction of the centroid shifts.

In total, we were able to extract the lifetimes of 30 excited states in the nucleus $^{96}$Ru. In eight cases, where lifetime values were known from previous measurements, our new values are in agreement. The large amount of new experimental data proves that this method is a powerful tool for lifetime measurements in the sub-picosecond regime, especially in cases where other techniques, such as the DSAM-INS technique, are not feasible.

\section*{Acknowledgements}
The authors would like to thank A.~Dewald for fruitful discussions as well as H.W. Becker and D. Rogalla from the Ruhr-Universit\"at Bochum for the assistance on the RBS measurement. We also highly acknowledge the support of the accelerator staff at the Institute for Nuclear Physics in Cologne during the experiment. This work is supported by the DFG under Grant No. ZI 510/4-2 and the Bulgarian Science Fund under Contract DFNI-E 01/2. S.G.P. and M.S. are supported by the Bonn-Cologne Graduate School of Physics and Astronomy.

\providecommand{\BIBCh}{Ch} \providecommand{\BIBZs}{Zs}
  \providecommand{\BIBGy}{Gy} \providecommand{\BIBTh}{Th}
  \providecommand{\BIBAn}{Analog Devices}

\end{document}